\documentclass[twocolumn,english,aps,prb,superscriptaddress,nopacs]{revtex4}
\usepackage{amsmath}
\usepackage{amssymb}
\usepackage{graphicx}
\usepackage{esint}

\makeatletter
\@ifundefined{textcolor}{}
{%
 \definecolor{BLACK}{gray}{0}
 \definecolor{WHITE}{gray}{1}
 \definecolor{RED}{rgb}{1,0,0}
 \definecolor{GREEN}{rgb}{0,1,0}
 \definecolor{BLUE}{rgb}{0,0,1}
 \definecolor{CYAN}{cmyk}{1,0,0,0}
 \definecolor{MAGENTA}{cmyk}{0,1,0,0}
 \definecolor{YELLOW}{cmyk}{0,0,1,0}
}

\renewcommand{\[}{\begin{equation}}

\renewcommand{\]}{\end{equation}}

\makeatother

\usepackage{babel}

\makeatother

\usepackage{babel}

\usepackage{sidecap}\sidecaptionvpos{figure}{c}

\makeatother

\usepackage{babel}

\makeatother

\usepackage{babel}

\makeatother

\usepackage{babel}
\begin{document}
\global\long\def\avg#1{\langle#1\rangle}

\global\long\def\p{\prime}

\global\long\def\dg{\dagger}

\global\long\def\ket#1{|#1\rangle}

\global\long\def\bra#1{\langle#1|}

\global\long\def\proj#1#2{|#1\rangle\langle#2|}

\global\long\def\inner#1#2{\langle#1|#2\rangle}

\global\long\def\tr{\mathrm{tr}}

\global\long\def\pd#1#2{\frac{\partial#1}{\partial#2}}

\global\long\def\spd#1#2{\frac{\partial^{2}#1}{\partial#2^{2}}}

\global\long\def\der#1#2{\frac{d#1}{d#2}}

\global\long\def\ders#1#2{\frac{d^{2}#1}{d#2^{2}}}

\global\long\def\im{\imath}

\title{Topological Schr\"odinger cats: Non-local quantum superpositions of topological defects}

\author{Jacek Dziarmaga}

\affiliation{Theoretical Division, MS-B213, Los Alamos National Laboratory, Los
Alamos, New Mexico 87545}

\affiliation{Institute of Physics and Centre for Complex Systems Research, Jagiellonian
University, Reymonta 4, 30-059 Krak\'ow, Poland}

\author{Wojciech H. Zurek}

\affiliation{Theoretical Division, MS-B213, Los Alamos National Laboratory, Los
Alamos, New Mexico 87545}

\author{Michael Zwolak}

\affiliation{Theoretical Division, MS-B213, Los Alamos National Laboratory, Los
Alamos, New Mexico 87545}

\affiliation{Department of Physics, Oregon State University, Corvallis, Oregon
97331}

\date{\today{}}

\maketitle
\textbf{Topological defects (such as monopoles, vortex lines, or domain
walls) mark locations where disparate choices of a broken symmetry
vacuum elsewhere in the system lead to irreconcilable differences
\cite{Mermin79-1,Michel80-1}. They are energetically costly (the
energy density in their core reaches that of the prior symmetric vacuum)
but topologically stable (the whole manifold would have to be rearranged
to get rid of the defect). We show how, in a paradigmatic model of
a quantum phase transition, a topological defect can be put in a non-local
superposition, so that \textendash{} in a region large compared to
the size of its core \textendash{} the order parameter of the system
is ``undecided'' by being in a quantum superposition of conflicting
choices of the broken symmetry. We dub such a topological Schr\"odinger
cat state a \textquoteleft{}Schr\"odinger kink\textquoteright{}, and
devise a version of a double-slit experiment suitable for topological
defects to describe one possible manifestation of the phenomenon.
Coherence detectable in such experiments will be suppressed as a consequence
of interaction with the environment. We analyze the environment-induced
decoherence and discuss its role in symmetry breaking.}

Topological defects are the epitome of locality. An example of a defect
occurs in the quantum Ising model where a lattice of spins interact
ferromagnetically with their nearest neighbors, i.e., the Hamiltonian
contains an interaction of the form $-\sigma_{n}^{z}\sigma_{n+1}^{z}$
for the $n^{th}$ spin on the lattice. The entire collection of spins
achieves its lowest energy when they are all aligned. However, there
are two choices for this lowest energy state, $|\cdots\uparrow\uparrow\uparrow\uparrow\cdots\rangle$
or $|\cdots\downarrow\downarrow\downarrow\downarrow\cdots\rangle$.
Both are energetically identical but each of them breaks the symmetry
of the Hamiltonian, which has no preference between ``up'' and ``down''.

\begin{figure*}[t]
\begin{centering}
\includegraphics[width=18cm]{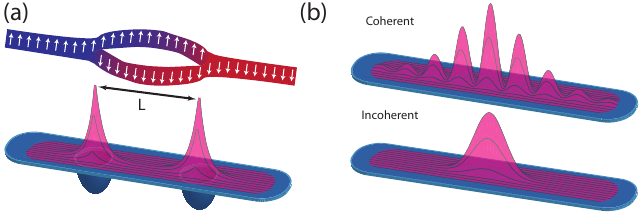} 
\par\end{centering}

\caption{A Schr\"odinger kink in a quantum Ising chain. (a) A topological defect
in a non-local superposition and (b) The analogue of a double-slit
experiment. A double-well potential (left) is used to create a superposition
of a topological defect, such as a Schr\"odinger kink described by wavefunction
representing a superposition of two locations (here we plot the corresponding
probability distribution in space). The analogy with the humane version
of the Schr\"odinger cat experiment is obvious. To carry out the double
slit experiment, the two potential wells are eliminated, allowing
the defect to move. In isolation, the two wavepackets emerging from
the ``slits'' interfere creating fringes of high and low probability
for the location of the kink. However, when the system interacts with
the environment, such a superposition will decohere at a rate proportional
to the distance $L$ \textendash{} the unzipped part of the chain
shown in (a), which corresponds to the size of the Schr\"odinger kink
\textendash{} resulting in a classical distribution for the defect.
\label{fig:Schematic}}
\end{figure*}

A configuration that includes topological defects could arise, for
example, in a driven phase transition, i.e., ``a quench''. In the
Ising model, a quench can be induced when the transverse field strength
$g$ is changed in the Hamiltonian 
\begin{equation}
H=-\sum_{n}\left(g\sigma_{n}^{x}+\sigma_{n}^{z}\sigma_{n+1}^{z}\right).\label{eq:Ising}
\end{equation}
 When the external field is strong enough ($g\gg1$), it ``wins''
and all spins align with $\sigma^{x}$. A decrease in $g$, however,
will lead to a phase transition at $g=1$ with the symmetry breaking
term $-\sigma_{n}^{z}\sigma_{n+1}^{z}$ trying to align all the spins
in their $z$-direction.

The choice of whether the spins will point up or down is made locally.
As a result of causality and the finite speed at which signals propagate,
the size of these domains will be determined by rate of change of
$g$ \cite{Kibble76-1,Kibble80-1,Zurek85-1,Zurek93-1,Zurek96-1,Kibble07-1}.
This will lead to configurations, such as $|\cdots\uparrow\uparrow\downarrow\downarrow\cdots\rangle$,
where the topological defect\textbf{ }marks the location of a ``switch''
from one broken symmetry ground state to the other. In this one-dimensional
example, the defect is a kink (a domain wall), but, in a sense, all
the topological defects (monopoles, vortex lines, and so on) that
exist in higher dimensions ``look the same''. A discussion of the
consequences of this process for the post-transition state is beyond
the scope of this work (but see, e.g., Refs. \onlinecite{Anglin99-1,Zurek05-1,Dziarmaga05-1}).
We only note that the density of the resulting kinks will be finite.

The quantum Ising model quenches a state according to Hamiltonian
\eqref{eq:Ising} so, when $g$ is changed in time, superpositions
of different locations of a kink (e.g., $\alpha|\cdots\uparrow\uparrow\downarrow\downarrow\downarrow\downarrow\cdots\rangle+\beta|\cdots\uparrow\uparrow\uparrow\downarrow\downarrow\downarrow\cdots\rangle+\gamma|\cdots\uparrow\uparrow\uparrow\uparrow\downarrow\downarrow\cdots\rangle$)
are allowed, and, indeed, inevitable \cite{Zurek05-1,Dziarmaga05-1}.
Spreading of a localized kink will come about as a consequence of
the kinetic term, $g\sigma_{n}^{x}$, in equation \eqref{eq:Ising}.
A convenient setting to investigate its effect is offered by a slight
modification of the original Hamiltonian, so that the Ising interaction
between a selected pair of sites, $n_{0},n_{0}+1$, is $-\left(1-w\right)\sigma_{n_{0}}^{z}\sigma_{n_{0}+1}^{z}$
with $w>0$, which differs somewhat from the uniform coupling of $-\sigma_{n}^{z}\sigma_{n+1}^{z}$.
This difference means that the kink is energetically less expensive
when localized between those two selected sites. The decrease in the
coupling constant by $w$ creates a local ``potential well'' that
binds this kink. On the other hand, the kinetic term will delocalize
the kink so that the quantum wavefunction of the kink will have the
form 
\begin{equation}
\psi_{n}\propto e^{-\gamma_{0}\left|n-n_{0}\right|},\label{eq:packet}
\end{equation}
 where $\psi_{n}$ is the amplitude for the kink to be on the link
$n$ between sites $n$ and $n+1$, $n_{0}$ is the location of the
potential well, and $\gamma_{0}=\sinh^{-1}\left(w/g\right)$ is the
inverse decay length of the wavepacket. One can imagine measurements
that will reveal such a non-local wavepacket \textendash{} a kink
in a superposition of many locations. We emphasize that the half-width
of this wavepacket is not the size of the kink: The kink is a local
object with a size given by the healing length that \textendash{}
in the quantum Ising model far away from the critical point \textendash{}
is given by the lattice spacing of the neighboring spins. The spread
in equation \eqref{eq:packet} represents a superposition of many
possible locations of the kink, which is bound to the weak link between
sites $n_{0}$ and $n_{0}+1$, but nevertheless has some spatial extent.

A tell-tale signature of quantum coherence is an interference pattern.
To see whether a defect can interfere with itself, we devise an analogue
of the double-slit experiment (see Figure \ref{fig:Schematic} for
a schematic). To this end, the local value of the Ising coupling can
be depressed by $w$ at two locations separated by $L$ sites. The
two links that bind the same kink initially are analogues of the slits
in the double-slit experiment. To achieve a situation where the kink
is ``both here and there'' one can start it on one of the two binding
sites and evolve so that tunneling of the kink will result in the
state 
\begin{align}
 & |\cdots\uparrow\uparrow\uparrow\overbrace{\downarrow\downarrow\cdots\downarrow\downarrow}^{L}\downarrow\downarrow\downarrow\cdots\rangle\nonumber \\
+ & |\cdots\uparrow\uparrow\uparrow\underbrace{\uparrow\uparrow\cdots\uparrow\uparrow}_{L}\downarrow\downarrow\downarrow\cdots\rangle,\label{eq:Super}
\end{align}
 where we have ignored the spread of the wavepacket for simplicity.
This is also illustrated in Figure \ref{fig:Schematic}(a), where
the superposition of the kink forms a ``scar'' in the orientation
of the spins, ``unzipping'' the $L$-sized region of the spin chain
in Fig. 1.

This is not the only way to create a ``Schr\"odinger kink'' state.
As we shall see later, one can also start with a kink localized state
on a single ``weak link''. If the kink is released symmetrically,
it will travel both left and right. %and then split this {}``weak link' to transport the kink adiabatically
%and symmetrically into a superposition of locations to the left and
%right. 
%In either case, retaining phase coherence between the two components of such Schr\"odinger kinks is crucial.

\begin{figure}[t]

\begin{centering}
\includegraphics[width=8cm]{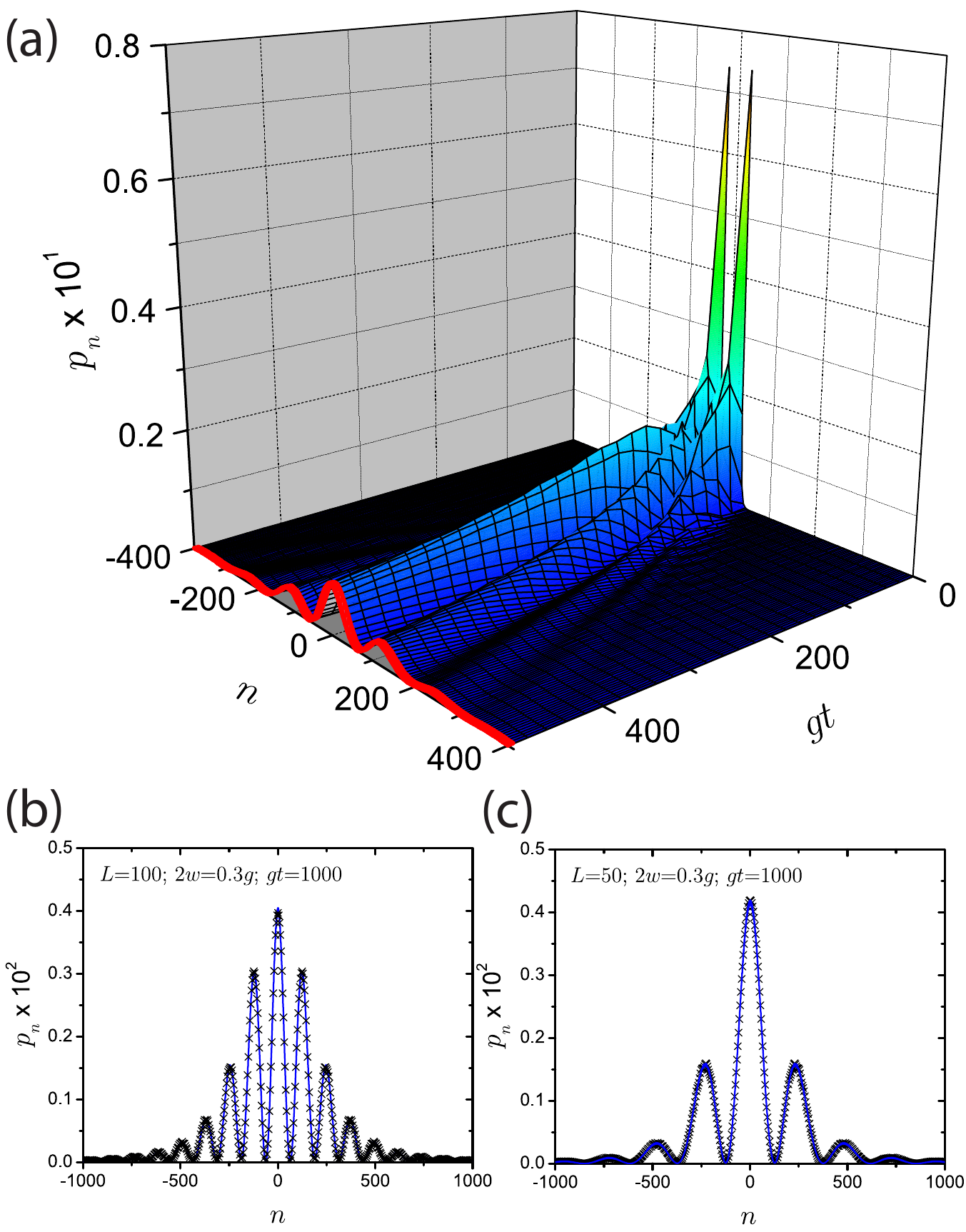} 
\par\end{centering}

\caption{Interference patterns after a Schr\"odinger kink is released. (a) Time
evolution of the Schr\"odinger kink for $L=50$ and $2w=0.3g$ results
in an interference pattern (highlighted in red at the final time).
(b) Interference pattern in the long time limit ($gt=1000$) for $L=100$
and $2w=0.3g$. (c) Interference pattern in the long time limit ($gt=1000$)
for $L=50$ and $2w=0.3g$. In (b) and (c), the exact data is shown
as black crosses and equation \eqref{eq:fringes} is shown as the
blue line. As the distance between the two ``slits'' where the kink
is trapped decreases so does the number of interference fringes. \label{fig:NoDecoherence}}
\end{figure}

The double-slit experiment can be now conducted starting from this
initial configuration by eliminating the two weak links. The kink
is then no longer bound by the ``double-well'' potential: The Schr\"odinger
kink propagates in accord with the Schr\"odinger equation and the two
components of the wavepacket run into each other. If coherence was
properly preserved between them, this will lead to an interference
pattern.

Preserving phase coherence is crucial if the resulting interference
pattern is to be seen in experiments. Trivial reasons for loss of
coherence \textendash{} such as an imprecise implementation of the
two weak links \textendash{} will have to be eliminated. However,
the fundamental reason for the loss of coherence is environment-induced
decoherence \cite{Zurek91-1,Paz00-1,Zurek03-1,Joos03-1,Schlosshauer08-1,Nielsen00-1}.
It is important to understand its causes and its nature, as it is
not just an impediment to creating superpositions described above,
but the prevailing reason why topological defects we encounter are
always localized.

Decoherence is caused by the interaction of individual spins with
the environment $\mathcal{E}$. The pointer states \cite{Zurek81-1,Zurek82-1}
entangle least with the environment. They are selected with the help
of the interaction Hamiltonian. For instance, an individual spin $\ket{\uparrow}$
can leave a different imprint on $\mathcal{E}$ than the spin $\ket{\downarrow}$,
i.e., $\left(\alpha\ket{\uparrow}+\beta\ket{\downarrow}\right)\ket{\mathcal{E}_{0}}\to\alpha\ket{\uparrow}\ket{\mathcal{E}_{\uparrow}}+\beta\ket{\downarrow}\ket{\mathcal{E}_{\downarrow}}$.
The overlap $\inner{\mathcal{E}_{\uparrow}}{\mathcal{E}_{\downarrow}}$
determines the decoherence factor, with $\inner{\mathcal{E}_{\uparrow}}{\mathcal{E}_{\downarrow}}=0$
corresponding to the complete loss of coherence. The decoherence factor
controls size of the off-diagonal terms in the density matrix. When
they disappear, quantum coherence between $\ket{\uparrow}$ and $\ket{\downarrow}$
is lost\cite{Zurek91-1,Paz00-1,Zurek03-1,Schlosshauer08-1}.

Returning to our Schr\"odinger kink, we note that when two locations
are separated by $L$ spins, equation \eqref{eq:Super}, the decoherence
process will take place simultaneously in all $L$ spins. Assuming
that each spin leaves its own imprint will lead to a decoherence factor
that scales as $\inner{\mathcal{E}_{\uparrow}}{\mathcal{E}_{\downarrow}}^{L}$,
where $L$ is the number of unzipped spins \textendash{} the spatial
extent of the superposition of the ``Schr\"odinger kink''. This exponential
scaling with the extent of the superposition is intuitively obvious:
We do not have to assume any specific model for the decoherence. All
that is needed is the familiar assumption (see, e.g., modeling of
errors in quantum error correction \cite{Nielsen00-1}) that individual
spins (or local regions) affect the environment individually.

This assumption suffices to show that the decoherence rate is proportional
to the ``size'', $L$, of the superposition. That is, decoherence
time is $\tau_{dec}\sim1/L$. This conclusion is confirmed by calculations
employing a master equation (see the Supplemental Information). One
can generalize this intuition to superpositions of topological defects
in higher dimensions by noting that it is the volume of the system
\textendash{} the size of the domain that is suspended in indecision
between two broken symmetry vacua \textendash{} that is responsible
for the decoherence rate.

\begin{figure}[t]
\begin{centering}
\includegraphics[width=8cm]{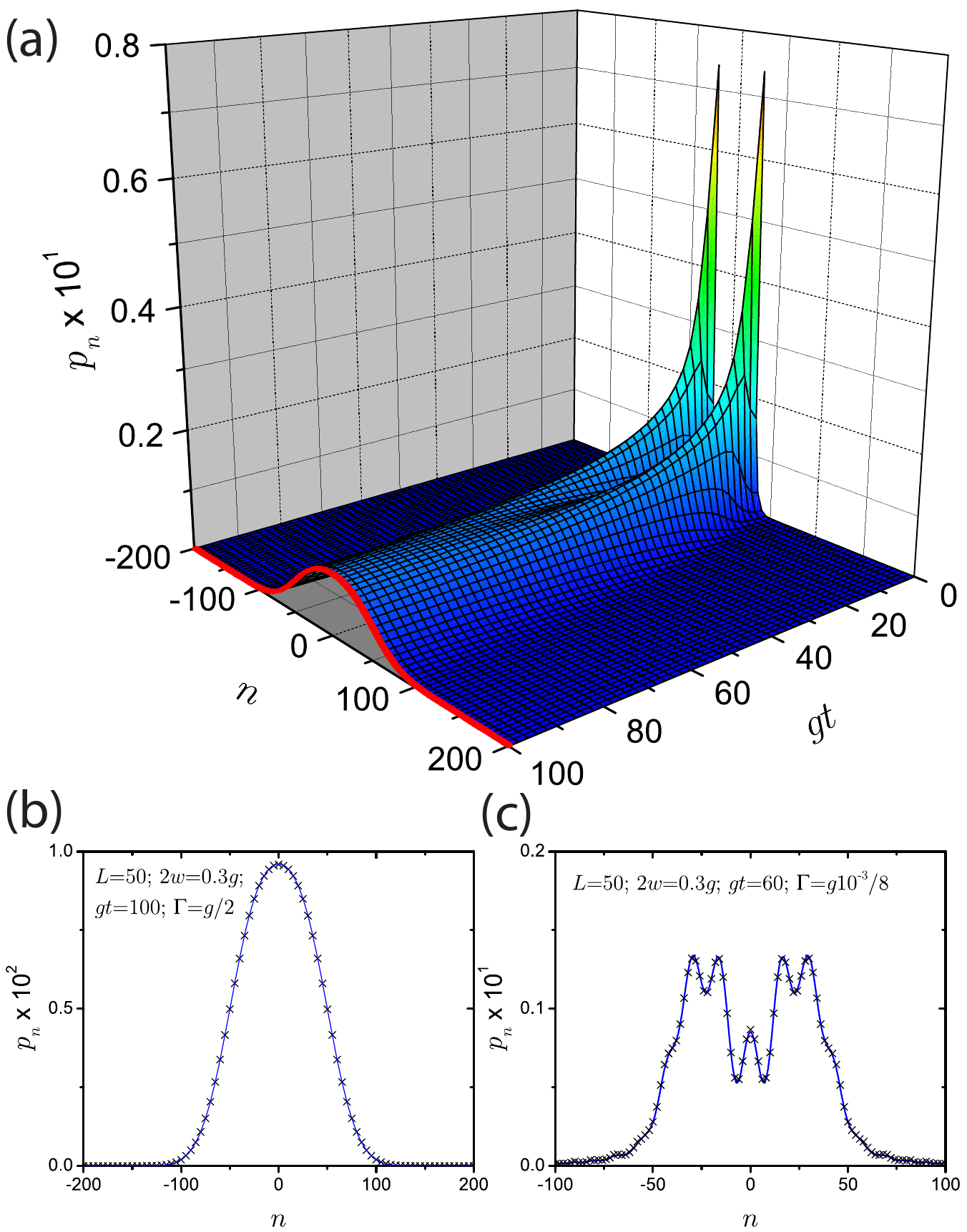} 
\par\end{centering}

\caption{A Schr\"odinger kink evolved in the presence of decoherence. (a) Time
evolution for strong decoherence ($\Gamma=g/2$), $L=50$, and $2w=0.3g$,
where the Schr\"odinger kink evolves into a (b) Gaussian mixture of
locations (highlighted in red in (a)) at a later time, $gt=100$ (the
black crosses are the exact data and the blue curve is the solution
to the diffusion equation). (c) Under weak decoherence ($\Gamma=g10^{-3}/8$),
the superposition is still visible at intermediate times, but the
decoherence smoothes out the fringes (the black crosses are the exact
data and the blue curve is the pure state solution convoluted with
a Lorentzian). \label{fig:Decoherence}}
\end{figure}

\begin{figure*}[t]
\begin{centering}
\includegraphics[width=18cm]{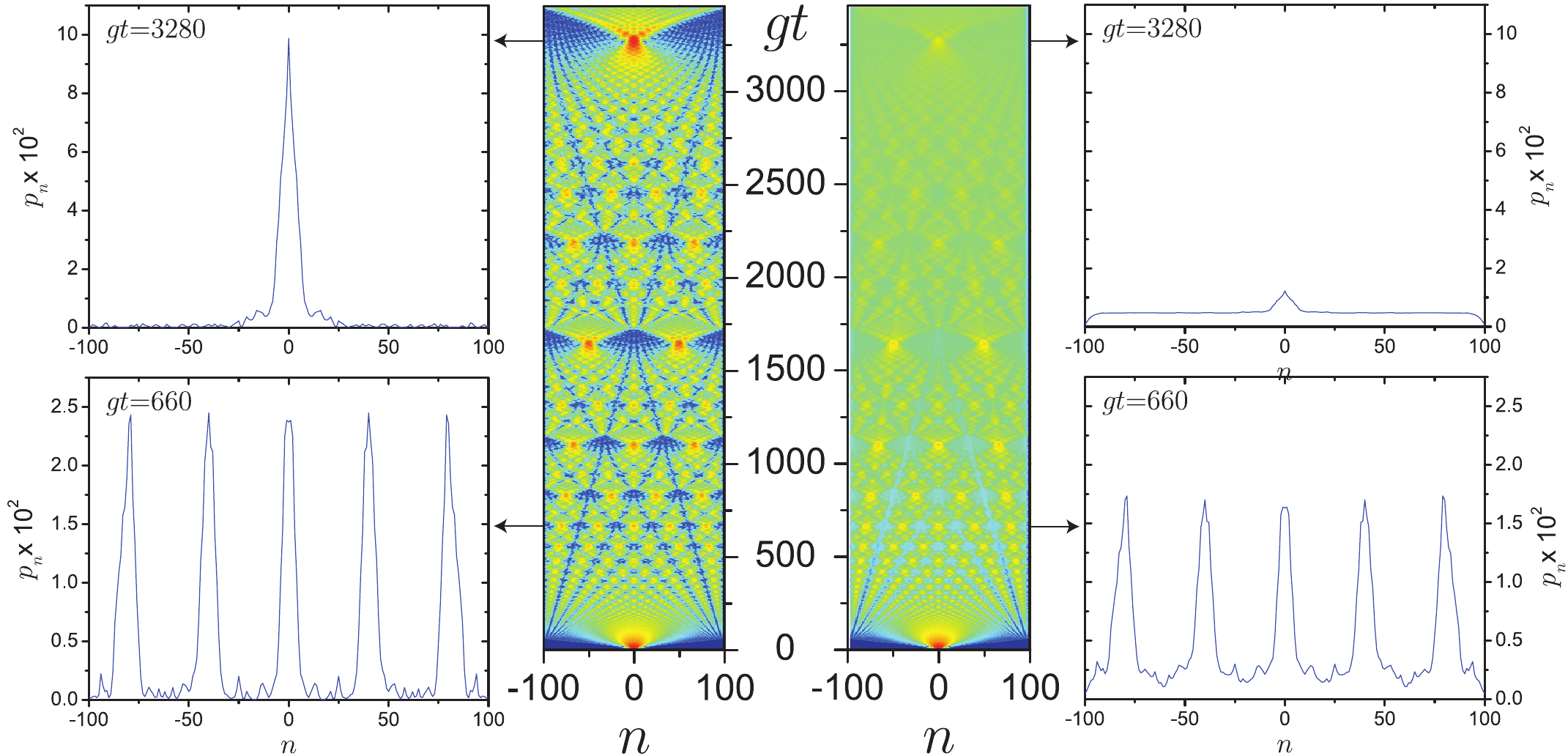} 
\par\end{centering}

\caption{A single kink evolved on a finite lattice with and without decoherence.
(Left) Time evolution without decoherence on an $L=201$ site lattice
and $2w=0.5g$. The kink travels outward on the lattice and after
reflecting off the boundaries, it starts to interfere with itself.
The leftmost panels show the self interference at two particular times.
(Right) The same simulation except in the presence of weak decoherence
($\Gamma=3g\times10^{-6}$). The kink can still interfere with itself,
but eventually decoherence will suppress the superposition, attenuating
the interference pattern. In the supplemental information, we show
movies of the development of this interference pattern for different
values of $w$. \label{fig:SelfInt}}
\end{figure*}

We now return to the Ising model. In the absence of decoherence, a
well-defined interference pattern develops when the two initial components
of the wavepacket run into each other (Fig. \ref{fig:NoDecoherence}).
As the distance, $L$, between these two components is increased,
more fringes become visible (i.e., decreasing $L$ decreases the magnitude
of the outer fringes), as seen by Figs. \ref{fig:NoDecoherence}(b,c).
This is demonstrated explicitly by the form of the fringes, 
\begin{equation}
p_{n}\left(t\right)=\left|\psi_{n}\left(t\right)\right|^{2}\propto\frac{1+\cos\frac{\left(n-n_{0}-L/2\right)L}{2gt}}{\left[1+\frac{\left(n-n_{0}-L/2\right)^{2}}{\left(2\gamma_{0}gt\right)^{2}}\right]^{2}}.\label{eq:fringes}
\end{equation}
 This interference pattern is analogous to the one observed in the
double-slit experiment. In the double-slit experiment the distance
between fringes is $\lambda D/L$, where $L$ is the distance between
the slits, $\lambda$ is the wavelength of light, and $D$ is the
distance to the screen. For kinks, the distance traveled is $D=\upsilon_{k}t\approx2gkt=2g\left(2\pi/\lambda\right)t$,
where $\upsilon_{k}=d\omega/dk\approx2gk$ is the group velocity of
the kink for slow kinks that are generated when $w\ll1$, as we assume
in deriving Eq. \eqref{eq:fringes} (the figures use $\upsilon_{k}=2g\sin k$,
demonstrating that this assumption is sound). This linear approximation
yields the fringe distance for kink interference $\lambda D/L=4\pi gt/L$,
as given in equation \eqref{eq:fringes}. The height of the second
peak relative to the first is $1/\left[1+4\pi^{2}/L^{2}\gamma_{0}^{2}\right]^{2}$.
In effect, the linear approximation of $\upsilon_{k}=2g\sin k$ entitles
one to think of kinks in a double slit experiment as one would think
of, say, electrons or neutrons.

Decoherence suppresses interference. Figure \ref{fig:Decoherence}(a,b)
shows the evolution of the kink probability density (that determines
the probability of finding a kink at a certain location) under strong
decoherence. In presence of decoherence it evolves into a Gaussian
form seen in Fig. \ref{fig:Decoherence}(b). Even weak decoherence
will attenuate interference. The decoherence time depends on $L$:
Coherence disappears exponentially fast in $L$, the separation between
components of the wavepacket. The evolution of the superposition under
weak decoherence is shown in Fig. \ref{fig:Decoherence}(c).

The double slit experiment for kinks we have just discussed has the
advantage of a straightforward interpretation. The use of the coherent
bi-local Schr\"odinger kink that is relatively weakly bound to the two
sites allows one to linearize the dispersion relation and use $\upsilon_{k}=2gk$
for the kink velocity. This leads in turn to the simple form of the
interference pattern, equation \eqref{eq:fringes}.

However, this ease of interpretation may come at the price of difficult
implementation. In particular, preparing the initial bi-local kink
wavepacket and maintaining coherence between its two pieces of the
will be a challenge. A different version of Schr\"odinger kink that
should be easier to implement is therefore illustrated in Fig. \ref{fig:SelfInt}.
Here the kink is initially bound to a specific link along the Ising
chain. This kink trap is instantaneously turned off, which results
in a coherent spreading of the wavepacket with a superposition of
velocities and in both directions on the Ising chain.

As before, we are not satisfied with just creating a Schr\"odinger kink.
One should confirm that quantum coherence is present. In Fig. \ref{fig:SelfInt}
this is accomplished by reflecting the spreading wavepacket from the
ends of the Ising chain. The time-evolving interference pattern is
now more complicated than before, but \textendash{} in absence of
decoherence \textendash{} it clearly exhibits quantum coherence that
can be predicted by suitably ``folding'' the kink's wavefunction
upon itself. Decoherence (as expected) suppresses interference fringes
over time: As with double slit analog, the decoherence strength will
have to decrease as $\sim1/L$, where $L$ is the support of the wavepacket,
if coherence \textendash{} and, hence, interference \textendash{}
is to survive. 

The interference pattern in Fig. 4 bears the imprint of the dispersion
relation on a lattice, $\omega=-2g\cos k$: When released, a tightly
bound kink turns into a wavepacket that propagates as a Bessel function,
$p_{n}\left(t\right)=\left|J_{n}\left(2gt\right)\right|^{2}$. Our
kinks of both Fig. 4 and especially of Fig. 2 are relatively weakly
bound. Therefore, the Bessel oscillation is suppressed \textendash{}
smoothed out by the non-local nature of the wavepacket that eliminates
large $k$ contributions (see Supplementary Information). Nevertheless,
small-scale jaggedness of the interference pattern visible in Fig.
4 (where $w$ is larger than in Fig. 2, and, therefore, the kink starts
more tightly localized) is a remnant of these Bessel oscillations.

An obvious application of this observation is the ``collapse'' of
the superposition of the broken symmetry vacua after a phase transition.
For example, in the case of the quantum Ising model, the ferromagnetic
ground state is a superposition of $\ket{\uparrow\cdots\uparrow}$
and $\ket{\downarrow\cdots\downarrow}$. The total number of spins,
$N$, in the Ising chain is the size of the superposition (e.g., $\ket{\uparrow\cdots\uparrow}+\ket{\downarrow\cdots\downarrow}$),
but this symmetric superposition will become a mixture of the two
obvious broken symmetry states in a very short time, $\tau_{dec}\sim1/N$.
This is a simple and compelling explanation of the symmetry breaking
that occurs whenever a phase transition starting from a symmetric
vacuum takes place.

The effectiveness of decoherence in localizing topological defects
provides novel insights into symmetry breaking dynamics. A phase transition
leads to a single quasi-classical configuration pockmarked with topological
defects: Decoherence is the final step in this process. It leads to
the ``collapse of the wavepacket'' that initially contains all the
possible broken-symmetry configurations.

In the end, only a simple quasi-classical configuration is in evidence.
This is essentially the same course of events that takes place in
quantum measurement \cite{Wheeler83-1}, where breaking of the unitary
symmetry allows for a ``collapse of the wavepacket''. Recent progress
in emulating quantum Ising and other models \cite{Friedenauer08-1,Kim10-1,Schneider10-1,Bakr10-1,Lin11-1,Chen11-1}
allows one to hope that experimental tests of ``Schr\"odinger kinks''
 may become possible in the near future. Moreover, although models
of superfluids do not allow one to develop and analyze microscopic
quantum theories of double-slit experiments for the relevant topological
defects with the detail we have presented above for kinks, experiments
involving the generation and manipulation of such defects in, for
example, gaseous Bose\textendash{}Einstein condensates allow one to
hope that testing of \textquoteleft{}topological Schr\"odinger cats\textquoteright{}
involving vortex lines or solitons may also be possible.
\begin{acknowledgments}
This research is supported by the U.S. Department of Energy through
the LANL/LDRD Program (W.H.Z. and M.Z.) and by the Polish Government
research project N202 124736 (J.D.). 
\end{acknowledgments}
\bibliographystyle{naturemag}
\bibliography{QCcorr,QPTsAndSCS}

\section*{Supplemental Information}

\subsection*{Interference fringes }

We are interested in the one-kink subspace of the Hamiltonian (1)
spanned by states 
\begin{equation}
\ket n=\ket{\uparrow\uparrow\cdots\uparrow_{n}\downarrow_{n+1}\cdots\downarrow\downarrow},
\end{equation}
 which have a kink on the link between sites $n,n+1$. We can confine
to this subspace when $g^{2}\ll1-w$ and the magnetic field $g$ is
too weak to mix with the subspaces of $3,5,..$ kinks. When there
is only one weaker link, then a stationary state $|\psi\rangle=\sum_{n}\psi_{n}\ket n$
satisfies a Schr\"odinger equation
\begin{equation}
E\psi_{n}=-g\left(\psi_{n+1}+\psi_{n-1}\right)-2w\delta_{n_{0},n}\psi_{n}.
\end{equation}
 The magnetic field $g$ provides a hopping term between nearest neighbor
links and the weaker link of strength $1-w$ results in a trapping
potential of strength $2w$ localized on the link $n_{0}$. This potential
has one localized bound state 
\begin{equation}
\psi_{n}=\cosh\left(\gamma_{0}\right)e^{-\gamma_{0}\left|n-n_{0}\right|},\label{eq:normpacket}
\end{equation}
 where $\gamma_{0}=\sinh^{-1}(w/g)$, with energy $E_{0}=-2g\cosh(\gamma_{0})~=~-2\sqrt{g^{2}+w^{2}}.$ 

When there are two weaker links, $n_{0}$ and $n_{0}+L$, the Schr\"odinger
equation 
\begin{equation}
E\psi_{n}=-g\left(\psi_{n+1}+\psi_{n-1}\right)-2w\left(\delta_{n_{0},n}+\delta_{n_{0}+L,n}\right)\psi_{n}
\end{equation}
 has two bound states
\begin{equation}
\psi_{n}^{\mp}\propto e^{-\gamma\left|n-n_{0}\right|}\mp e^{-\gamma\left|n-n_{0}-L\right|}\label{eq:mp}
\end{equation}
 with energy $E=-2g\cosh(\gamma)$ and $\gamma$ being one of the
two solutions of the equation 
\begin{equation}
\left[1-\frac{g\sinh\left(\gamma\right)}{w}\right]^{2}=e^{-2\gamma L}.
\end{equation}
 The greater of the two $\gamma$'s corresponds to the symmetric ground
state $\psi^{+}$.

We are interested in the ``tight-binding'' regime where the wave
packets in equation \eqref{eq:mp} overlap weakly, i.e., $e^{-\gamma L}\ll1$.
In this regime $\gamma\approx\gamma_{0}$ in both bound states $\psi^{\mp}$,
their energies are $E\approx E_{0}\pm\omega$, respectively, where
the gap
\begin{equation}
2\omega=\frac{4w^{2}}{\sqrt{g^{2}+w^{2}}}e^{-2\gamma_{0}L}
\end{equation}
 is relatively small, $2\omega\ll E_{0}$. In this regime, we can
initially prepare the ground state of a single well, equation \eqref{eq:normpacket},
which, in the basis \eqref{eq:mp}, reads $\psi^{+}+\psi^{-}$. After
this preparation, we can either switch on the second well suddenly
or turn it on adiabatically. An alternative preparation is to split
a single well adiabatically and symmetrically into two wells. With
a real-time preparation, the initial state evolves into $\psi_{n}^{+}e^{+i\omega t}+\psi_{n}^{-}e^{-i\omega t}~$
and the state becomes an equal superposition of the left and right
wells, $e^{-\gamma_{0}|n-n_{0}|}~+~i~e^{-\gamma_{0}|n-n_{0}-L|}~,$
at the earliest time $\omega t=\frac{1}{4}\pi$. In the adiabatic
case, the preparation ends in the ground state $\psi^{+}$ of the
double well. Since both cases require roughly the same time $\simeq1/\omega$,
we opt for the more robust adiabatic preparation of $\psi^{+}$.

Once the state $\psi^{+}$ has been prepared, we switch off the double-well
potential at $t=0$, $w\to0$, to let the wave function freely disperse
with just the hopping term,
\begin{equation}
i\der{}t\psi_{n}=-g\left(\psi_{n+1}+\psi_{n-1}\right).\label{eq:hopping}
\end{equation}
 When $2g\gamma_{0}^{2}~t~\gg~1$, the probability distribution develops
an interference pattern 
\begin{equation}
p_{n}\left(t\right)=\left|\psi_{n}\left(t\right)\right|^{2}\propto\frac{1+\cos\frac{\left(n-n_{0}-L/2\right)L}{2gt}}{\left[1+\frac{\left(n-n_{0}-L/2\right)^{2}}{\left(2\gamma_{0}gt\right)^{2}}\right]^{2}}.\label{eq:fringesSI}
\end{equation}
 Here the distance between fringes is $4\pi gt/L$ and the width of
the Lorentzian-squared envelope is $2\gamma_{0}gt$. In the tight-binding
regime, where $\gamma_{0}L\gg1$, we obtain a large number of fringes.
Equation \eqref{eq:fringesSI} is the result we plot with the exact
solution within the main text. 

If, on the other hand, one starts from a single kink localized on
link 0, $\psi_{n}(t=0)=\delta_{n,0}$, it will evolve into 
\[
\psi_{n}(t)\sim\int_{-\pi}^{\pi}dke^{2igt\cos k}e^{ikn}\sim J_{n}(2gt),
\]
 where $J_{n}$ is a Bessel function. This will also give interference,
but not of the form in equation \eqref{eq:fringesSI}. We are interested
in the regime where these ``intrinsic'' oscillations due to the
lattice are negligible. That is, if we start with a kink wavepacket
with some spread, $\gamma_{0}$ from above, then
\[
\psi_{n}(t)\sim\int_{-\pi}^{\pi}dke^{2igt\cos k}e^{ikn}f[k/\gamma_{0}],
\]
where $f$ will give the Fourier transform of the initial wavepacket
that cuts off large $k$ compared to $\gamma_{0}$. When the kink
is initially tightly bound tightly (large $w$ and $\gamma_{0}$),
then $f$ is a constant for $-\pi<k<\pi$ and we obtain $\sim J_{n}(2gt)$.
When $w$ is weak and $\gamma_{0}\ll1$, then $f$ cuts off $\left|k\right|$
greater than $\gamma_{0}$ and we can approximate $\cos k\approx1-k^{2}/2$,
i.e., we can linearize the group velocity and obtain equation \eqref{eq:fringesSI}.
The three movies show the development of the interference pattern
(of Fig. 4) for different values of $w$.

\subsection*{Decoherence}

Once the double-well potential has been switched off there are no
energy gaps $2\omega$ or $E_{0}$ to protect against even weak decoherence.
Under local, Markovian dephasing, the state, described by a density
matrix $\rho\left(t\right)$, evolves according to a master equation
\begin{equation}
\der{}t\rho\left(t\right)=-i\left[H_{0},\rho\left(t\right)\right]-\frac{\Gamma}{4}\sum_{n}\left[\sigma_{n}^{z},\left[\sigma_{n}^{z},\rho\left(t\right)\right]\right],
\end{equation}
 where $H_{0}$ is the hopping Hamiltonian that gives rise to the
evolution in equation \eqref{eq:hopping} and the initial state is
$\rho(0)=|\psi^{+}\rangle\langle\psi^{+}|$. In the position representation,
$\rho=\sum_{m,n}\rho_{m,n}|m\rangle\langle n|$, the master equation
reads
\begin{align}
\der{\rho_{m,n}}t= & ig\left(\rho_{m+1,n}+\rho_{m-1,n}-\rho_{m,n+1}-\rho_{m,n-1}\right)\nonumber \\
 & -\Gamma\left|m-n\right|\rho_{m,n}.\label{eq:mastermn}
\end{align}
 We can consider regimes of strong or weak decoherence when either
$\Gamma\gg g$ or $\Gamma\ll g$ respectively.

\subsubsection*{Strong decoherence}

When we temporarily set $g=0$, the off-diagonal matrix elements decay
like $\rho_{m,n}(t)=\rho(0)\exp(-\Gamma|m-n|t)$. Therefore, we can
assume that deep in this regime only the diagonal elements, $\rho_{n,n}\equiv p_{n}$,
and near-diagonal elements, $\rho_{n,n+1}=z_{n}$ and $\rho_{n+1,n}=z_{n}^{*}$,
are non-zero. All other elements, even if they are non-zero initially,
quickly become negligible. The master equation then simplifies to
the set of equations 
\begin{equation}
\der{p_{n}}t=g\left(s_{n-1}-s_{n}\right)
\end{equation}
 and
\begin{equation}
\der{s_{n}}t=-\Gamma s_{n}-2g\left(p_{n+1}-p_{n}\right),
\end{equation}
 where $s_{n}=-i\left(z_{n}-z_{n}^{*}\right)$. Since $\Gamma\gg g$,
the slave field $s_{n}$ can be adiabatically eliminated and we obtain
a lattice diffusion equation
\begin{equation}
\ders{p_{n}}t=D\left(p_{n+1}-2p_{n}+p_{n-1}\right)
\end{equation}
 with the diffusion constant
\begin{equation}
D=\frac{2g^{2}}{\Gamma}.
\end{equation}
 When $2Dt\gg1$, the initial distribution $p_{n}=|\psi_{n}^{+}|^{2}$
spreads into 
\begin{equation}
p_{n}^{(\Gamma\gg g)}(t)=\sum_{m}|\psi_{m}^{+}|^{2}\frac{1}{\sqrt{4\pi Dt}}e^{-\frac{\left(n-m\right)^{2}}{4Dt}}\label{eq:Gaussian}
\end{equation}
 without any interference fringes.

\subsubsection*{Weak decoherence}

In case of weak decoherence, $\Gamma\ll g$, the initially smooth
wavefunction does not get localized in space so we can make a long
wavelength approximation (LWA) where the lattice site numbers $m,n$
are continuous coordinates. In quasimomentum representation, 
\[
\rho_{p,q}(t)=\sum_{m,n}\rho_{m,n}\exp[i(pm-qn)],
\]
 after going to the interaction picture and making the LWA, 
\begin{eqnarray}
\rho_{p,q}(t) & = & \tilde{\rho}_{p,q}(t)\exp[2igt(\cos p-\cos q)]\nonumber \\
 & \approx & \tilde{\rho}_{p,q}(t)\exp[-igt(p^{2}-q^{2})],
\end{eqnarray}
 the master equation \eqref{eq:mastermn} becomes 
\[
\frac{d\tilde{\rho}_{p,q}}{dt}=-\Gamma\int_{-\infty}^{\infty}\frac{dk}{2\pi}~e^{-2igt(p-q)k}~\tilde{\rho}_{p+k,q+k}~\int_{-\infty}^{\infty}dm~|m|e^{-ikm}~.
\]
 Its exact solution gives a probability distribution in space 
\begin{equation}
p_{n}^{(\Gamma\ll g)}(t)~=~\frac{1}{\pi}~\sum_{m}\frac{l(t)}{l^{2}(t)+m^{2}}~p_{n+m}(t),\label{eq:LConv}
\end{equation}
 where $l(t)=g\Gamma t^{2}$ and $p_{m}(t)$ is a probability distribution
in the absence of decoherence, i.e., the interference fringes in equation
(4).

The Lorentzian convolution \eqref{eq:LConv} coarse-grains the fringes
in $p_{m}(t)$ on the scale $l(t)$. $l(t)$ becomes greater than
the distance between fringes at $t_{{\rm dec}}~\simeq~4\pi/\Gamma L$.
This is the (not quite unexpected) decoherence time when the environment
can distinguish between the $\uparrow$ and $\downarrow$ magnetization
of the $L$ sites between the potential wells.

The convolution \eqref{eq:LConv} tends to a Lorentzian when $l(t)$
is much greater than the width of the envelope in equation (4) or,
equivalently, $\Gamma t\gg\frac{1}{2}\gamma_{0}$. The Lorentzian
is different (wider) than the Lorentzian-squared envelope in the decoherence-free
fringes (4) and the asymptotic Gaussian \eqref{eq:Gaussian} in the
strong decoherence limit.
\end{document}